\documentclass[aps,pra]{revtex4-1}
\usepackage{blindtext}

\usepackage{graphicx}
\usepackage{amssymb}
\usepackage{amsmath}
\usepackage{bm}
\usepackage{braket}
\usepackage{gensymb}
\usepackage{footmisc}
\usepackage{booktabs}
\usepackage{array}
\usepackage{hyperref}
\usepackage{xcolor}
\hypersetup{
    colorlinks,
    linkcolor={black},
    citecolor={blue},
    urlcolor={blue}
}

\newcommand{\RomanNumeralCaps}[1]
    {\MakeUppercase{\romannumeral #1}}

\begin{document}

\title{Non-Abelian gauge potential driven localization transition in quasiperiodic optical lattices}

\author{En Guo Guan, Hang Yu, and Gang Wang}
\email{phwanggang@gmail.com}

\affiliation{ School of Physical Science and Technology, Soochow
University, Suzhou 215006, China }


\begin{abstract}
Gauge potential is an emergent concept in systems of ultracold atomic
gases. Derived from quantum waves immersed in an \emph{Abelian}
gauge, the quasiperiodic Aubry-Andre-Harper (AAH) model is a simple
yet powerful Hamiltonian to study the Anderson localization of
ultracold atoms. In this work, we investigate the localization
properties of ultracold atoms trapped in quasiperiodic optical
lattices subject to a non-Abelian gauge, which can be depicted by a
family of non-Abelian AAH models. We identify that the non-Abelian
AAH models can bear the self-duality under the Fourier
transformation. We thus analyze the localization transition of this
self-dual non-Abelian quasiperiodic optical lattices, revealing that
the non-Abelian gauge involved drives a transition from a pure
delocalization phase, then to coexistence phases, and then finally to
a pure localization phase. This is in stark contrast to the Abelian
AAH model that does not support the coexistence phases.
Our results thus comprise a new insight on the fundamental aspects of
Anderson localization in quasiperiodic systems, from the perspective
of non-Abelian gauge.
\end{abstract}

\maketitle


\section{Introduction}

A magnetic field, through its geometric gauge potential, usually
causes measurable effects on the wave functions. This concept of
gauge potential has been fueling research since Aharonov and Bohm's
seminal work~\cite{AB}. In particular, the recent explorations of
gauge fields synthesized for neutral atoms open up many avenues in
the field of ultracold atomic
gases~\cite{JakschNJP2003,NonAbelian:20051,NonAbelian:20052,DalibardRMP2011,GoldmanRPP2014,AidelsburgerCRP2018,GalitskiPhysTod2019}.
Among them the synthetic non-Abelian gauge in the atomic gases with
internal degrees of freedom is of special interest because of the
high gauge symmetry. The non-Abelian gauge will emerge in cold-atom
systems when the orbital motion is coupled to the internal hyperfine
levels. Therefore, the non-Abelian gauge is usually responsible for
various (pseudo)spin-orbit interactions in the ultracold atoms. In
this case, the Aharonov-Bohm phase will be replaced by a matrix that
can bear its imprint on the different internal
states~\cite{WilczekPRL1984}.

In parallel to the synthetic gauge potentials, another timely topic
in the ultracold atoms is the Anderson localization, where
researchers try to understand the intricate effects of disorder. The
effect of disorder originates from the interference of multiple
scattering paths in the disordered configurations. Therefore, strong
enough disorder could bring about the destructive interference, so
that arrest transport completely, leading to a phase transition from
delocalized (``metal" phase) to exponentially localized (``insulator"
phase)
states~\cite{AndersonPR1958,GangofFour,LeeRMP1985,Kramer1993,MirlinRMP2008}.
The concept of Anderson localization has been progressively developed
from its original scope of solid-state physics to the context of
ultracold atomic gases, in which a vast number of model Hamiltonians
can be realized by engineering appropriate optical
potentials~\cite{BlochRMP2008}. Within this setting, a range of
fascinating localization phenomena has been revealed successfully,
including the direct observations of weak
localization~\cite{CBS1,CBS2}, the Anderson localization in
1D~\cite{BillyNature2008,RoatiNature2008} and 3D
systems~\cite{3DAL1,3DAL2}, and mobility edges~\cite{me1}, just to
mention a few.

Concerning the Anderson localization of ultracold atoms,
quasiperiodic optical lattices arouse a recent surge of
interest~\cite{RoatiNature2008}. Aubry and Andr\'{e} have
analytically demonstrated that the one-dimensional lattice with a
quasiperiodic modulation, the so-called Aubry-Andr\'{e}-Harper (AAH)
model~\cite{AA,Harper}, can show the localization transition. One of
the key features of the AAH model is either \emph{all} states being
extended or localized, depending on the modulation strength of the
quasiperiodic potential. This localization transition in the space of
modulation strength arises from the self-duality of the AAH model. As
a result, either ballistic or localized excitations were observed in
the AAH optical lattices~\cite{RoatiNature2008}. Distinct from the
truly random disorder, the disorder for the quasiperiodic lattice is
introduced in a deterministic manner. Even so, the coexistence phase,
defined as the regime in which the localized and extended states
exist simultaneously under the same disorder strength, can appear in
1D quasiperiodic
models~\cite{Sarma88,Holthaus2007,Thouless88,Biddle10,SF1,SF2,SF3,SF4,selfdual4,sarma4,sarma5},
as is the case for the higher-dimensional disordered models with true
randomness. The coexistence phase has been experimentally observed in
the quasiperiodic optical lattices~\cite{me2,me3}.

Motivated by the ongoing progress in synthetic gauge potential and
Anderson localization using ultracold atoms, several natural while
interesting questions arise, from the point of view of the
non-Abelian gauge: Under the presence the non-Abelian gauge is it
possible to construct a family of self-dual models? If that, what is
the localization property of such self-dual non-Abelian lattices? And
what is the associated phase diagram?

In the current work we answer these questions through studying the
cold atoms trapped in quasiperiodic optical lattices which are
subject to non-Abelian SU(2) gauge potentials. The optical lattices
can be modeled by a family of non-Abelian AAH Hamiltonians. We show
that the non-Abelian gauge has dramatic consequences on the
localization of ultracold atoms in
quasiperiodic optical lattices. To be specific, 
such non-Abelian quasiperiodic lattices become self-dual under the
Fourier transformation, whereas the coexistence phase composed of
localized and extended states can emerge, manifesting itself as
definite mobility edges in the energy space. We present evidence that
supports these observations by employing two kind of measures of
localization designed to treat the quasiperiodic lattices, inclusive
of inverse participation ratio and the decomposition of spectrum.
Comparing to its Abelian counterpart, we reveal a rich phase diagram
driven by the non-Abelian gauge, featuring a transition of
metal-coexistence \RomanNumeralCaps{1}-coexistence
\RomanNumeralCaps{2}-Anderson insulator. This is in marked contrast
with the Abelian AAH model, therein being self-dual while having no
coexistence phase.

The paper is organized as follows. In Sec.~\ref{sec:abelianAAH}, we
firstly review the results of Anderson localization in 1D
quasiperiodic Abelian AAH model, which will be useful in the
following. Then, we formulate the non-Abelian optical lattices with
quasiperiodic modulations in Sec.~\ref{sec2}, and analyze the
relevant localization properties. We finally summarize our results in
Sec.~\ref{sec:conclusions}. Hereafter, the non-Abelian AAH model will
be referred to as the NA-AAH model so as to distinguish it from its
Abelian counterpart, i.e., the original AAH model (abbreviated as
A-AAH model).

\section{Short review of quasiperiodic A-AAH Hamiltonian} \label{sec:abelianAAH}

Before elaborating on our results, it will be instructive to make
some general reviews on the localization properties of the standard
quasiperiodic A-AAH lattice~\cite{AA}. The parent model for the
one-dimensional A-AAH lattice is the Hofstadter Hamiltonian,
depicting the electrons hopping in the square lattice threaded by a
homogeneous magnetic field corresponding to an \emph{Abelian} $U(1)$
gauge potential. The Hofstadter Hamiltonian reads
\begin{equation}
\mathcal{H}=-\sum_{m,n} J_x \psi_{m+1,n}^{\dagger} \psi_{m,n}
+ J_y \psi_{m,n+1}^{\dagger} e^{i \theta_y} \psi_{m,n}+\textit{h.c.}
\label{eq:Hofstadter_model}
\end{equation}
under the Landau gauge~\cite{HofstadterPRB}. Here
$\psi_{m,n}^\dagger$, $\psi_{m,n}$ are the creation and annihilation
operators on the lattice site $(x=m$, $y=n)$, the phase factors are
given by $\theta_y=2\pi\Omega m$ where $\Omega$ 
is the magnetic flux $\Phi$ in units of the
flux quantum $\Phi_0$ penetrating each plaquette, 
$J_{x,y}$ represents the nearest-neighbor hopping amplitudes. In a
mixed momentum-position space as obtained by performing the Fourier
transformation only along the $y$ direction,
Eq.~\ref{eq:Hofstadter_model} reduces to the celebrated A-AAH
Hamiltonian~\cite{AA}
\begin{equation}
\mathcal{H}=-\sum_{m} \,
\psi_{m+1}^\dagger \psi_{m} + \psi_{m-1}^\dagger \psi_{m}
+ 2 \lambda \cos(2\pi \Omega m + k_y)\, \psi_{m}^\dagger \psi_{m},
\label{eq:AAH_model}
\end{equation}
where $\lambda$ parametrizes the ratio $J_y/J_x$ and denotes the
modulation strength of the potential. An irrational $\Omega$ (usually
the golden mean) ensures the quasiperiodic lattices. Without loss of
generality, we assume $k_y=0$ below. One premier feature of the
quasiperiodic A-AAH model is its self-duality: Upon Fourier
transformation, which exchanges the real and momentum space,
Eq.~(\ref{eq:AAH_model}) can be proven to be dual to itself, with
$\lambda \rightarrow \lambda^{-1}$. Exactly at $\lambda=1$ the model
is invariant under the duality, identified as the transition point to
separate the pure localization and pure delocalization phases.
Therefore, for $\lambda < 1$ all the states are delocalized, while
$\lambda > 1$ makes all of the states exponentially localized. This
model shows the localization-delocalization transition without the
presence of the energy-dependent mobility edges in the phase diagram.
Such a transition has been observed experimentally in a
noninteracting bosonic gas~\cite{RoatiNature2008}. On the other hand,
there are variations that enrich the localization
transition~\cite{Sarma88,Holthaus2007,Thouless88,Biddle10,SF1,SF2,SF3,SF4,selfdual4,sarma4,sarma5,me2,me3},
e.g., the emergence of mobility edges in the 1-dimensional
quasiperiodic systems as that in 3-dimensional space of truly random
disorder.

\begin{figure}[htbp]
\centering
\includegraphics[width=0.7\linewidth]{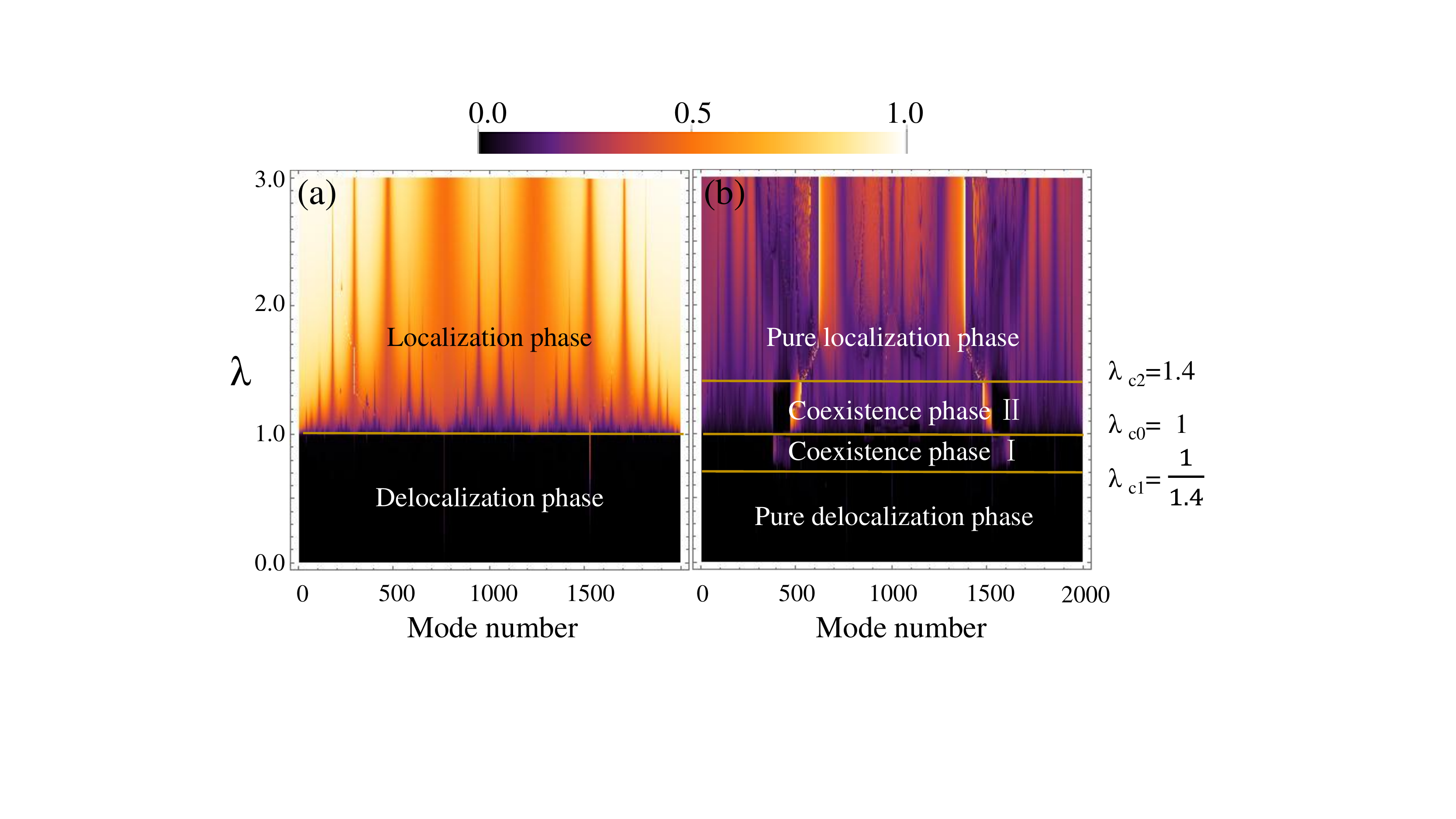}
\caption{(Color online) The IPR distribution of states as a function of $\lambda$.
(a)A-AAH model, (b) NA-AAH model with a fixed non-Abelian strength $q=0.3\pi$.
The states are sorted in descending order according to
their energies. The lines are indicative of the critical modulation strengths to separate
various phases, and the relevant values are estimated through spectral measures elaborated
below in subsec.~\ref{subsec:localization in NA-AAH}.}
\label{fig2:IPR}
\end{figure}

To quantify the localization properties of the wave functions, we
calculate the inverse participation ratio (IPR) of states defined as
follows~\cite{Kramer1993}:
\begin{equation}
\mathrm{IPR}^{(i)}=\frac{\sum_m|\psi_m^{(i)}|^4}{(\sum_m|\psi_m^{(i)}|^2)^2},
\end{equation}
where the superscript $i$ denotes the $i$th state $\psi^{(i)}$. For a
spatially extended state, the IPR approaches zero whereas it is
finite for a localized state. Therefore, the IPR can be taken as a
criterion to distinguish the extended states from the localized
ones~\cite{Kramer1993}. Figure~\ref{fig2:IPR} (a) illustrates clear
distinctions of IPRs for various modulation strength. Obviously this
indicates that a sharp localization-delocalization transition occurs
in the quasiperiodic A-AAH model at the dual point $\lambda=1$,
beyond which \emph{all} states convert from extended to localized.

\section{Anderson localization in quasiperiodic NA-AAH optical lattices}\label{sec2}

\subsection{NA-AAH Hamiltonian and its self-duality} \label{sec:non-Abelian AAH
model}

The gauge field used in the original AAH model is of the Abelian
type. Quite recently, there are proposals to generate non-Abelian
gauge fields in optical
lattices~\cite{NonAbelian:20051,NonAbelian:20052}. Now we turn to the
quasiperiodic non-Abelian AAH model, which originates from the
Hofstadter Hamiltonian in a non-Abelian optical
lattice~\cite{GoldmanPRL2009}, and investigate its localization
properties.

Our starting point is the two-component cold atoms characterized by
two internal degrees of freedom, providing a (pseudo)spinor
$\Psi=(\psi^\uparrow,\psi^\downarrow)^T$. The ultracold gases are
trapped in a 2-dimensional optical square lattice. We impose a
synthetic non-Abelian gauge potential on the optical lattice
\begin{equation}
\mathbf{A}= \bigl ( q  \sigma_y , 2 \pi \Omega m \sigma_0 +q \sigma_x , 0 \bigr ).
\label{gauge}
\end{equation}
Here $\sigma_{x(y)}$ denotes the Pauli operator. The gauge field
$\mathbf{A}$ is constituted by an Abelian U(1) term $\mathbf{A}_{a}=
\bigl (0, 2\pi \Omega m \sigma_0 , 0 \bigr )$, and by a constant
SU(2) term $\mathbf{A}_{na}= q\bigl (\sigma_y, \sigma_x, 0 \bigr)$
proportional to the parameter $q$, quantifying the strength of the
non-Abelian gauge. This gauge potential may be generated using a two
dimensional optical superlattice based on laser-assisted
spin-dependent hopping
~\cite{NonAbelian:20051,NonAbelian:20052,AHE2,AHE3,AE,GoldmanPRL2009}.
The non-Abelian gauge flux $q$ is determined by the Rabi frequencies
of the two-photon off-resonant transition and can be tuned via
changing the laser field intensity. The $\mathrm{SU(2)}$ non-Abelian
part is the new ingredient as compared with the original Hofstadter
Hamiltonian (Eq.~\ref{eq:Hofstadter_model}). According to the Peierls
substitution, the non-Abelian component imprints an
internal-state-dependent phase factor on the hoppings, and changes
the internal states of the spinor. The corresponding non-Abelian
Hofstadter Hamiltonian reads~\cite{GoldmanPRL2009}
\begin{equation}
H=-\sum\limits_{m,n} \Psi_{m+1,n}^{\dagger } J_x e^{i q \sigma
_{y}}\Psi_{m,n}+\Psi_{m,n+1}^{\dagger } J_y e^{i\left( q \sigma _{x}+2\pi \Omega
m\right) }\Psi_{m,n} +h.c..
\label{eqn:NA_Hofstadter_ham}
\end{equation}
By taking periodic and open boundary conditions along $y$ and $x$
directions separatively, the Hamiltonian will be transformed into
\begin{align}
&\mathcal{H} = -\sum\limits_{m} 2 \lambda \Psi_m^{\dagger} \begin{pmatrix}
\cos q \cos (2 \pi \Omega m )  &  -\sin q \sin(2 \pi \Omega m ) \\
-\sin q \sin (2 \pi \Omega m ) &  \cos q \cos (2 \pi \Omega m ) \end{pmatrix} \Psi_m \notag \\
&+ \Psi_{m+1}^{\dagger} \begin{pmatrix} \cos q  & \sin q \\
-\sin q & \cos q  \end{pmatrix} \Psi_{m} + \Psi_{m-1}^{\dagger} \begin{pmatrix} \cos q  & -\sin q \\
\sin q & \cos q  \end{pmatrix} \Psi_{m}.
\label{eq:nonabelianharper}
\end{align}
Compared with Eq.~\ref{eq:AAH_model}, the Hamiltonian
Eq.~\ref{eq:nonabelianharper} involves a non-Abelian gauge. Thus we
name it non-Abelian AAH model.

The Fourier transformation of Eq.~\ref{eq:nonabelianharper} provides
the following equation in the dual space:
\begin{align}
&\mathcal{H}^{\ast} =-\sum\limits_{k} 2 \Psi_k^{\dagger} \begin{pmatrix}
\cos q \cos (2 \pi \Omega k )  &  -\sin q \sin(2 \pi \Omega k ) \\
-\sin q \sin (2 \pi \Omega k ) &  \cos q \cos (2 \pi \Omega k )  \end{pmatrix} \Psi_k \nonumber \\
&+ \lambda \Psi_{k+1}^{\dagger} \begin{pmatrix} \cos q  & \sin q \\
-\sin q & \cos q  \end{pmatrix} \Psi_{k} +
\lambda \Psi_{k-1}^{\dagger} \begin{pmatrix}
\cos q  & -\sin q \\ \sin q & \cos q \end{pmatrix} \Psi_{k}.\label{eq:nonabelianharper-in-k}
\end{align}
Definitely, Eq.~\ref{eq:nonabelianharper-in-k} takes the same form as
that in real space (Eq.~\ref{eq:nonabelianharper}) with just the
coefficients being interchanged. Hence the NA-AAH model embodies the
self-duality, ensuring that a state which is extended in the
parameter space $(\lambda,~q)$ will be converted into a localized
state in $(\frac{1}{\lambda},~q)$, and vice versa. As a result,
$\lambda_{c0}=1$ is identified as a transition point.

\subsection{localization properties}\label{subsec:localization in NA-AAH}

Having described the self-duality of the NA-AAH optical lattices, we
now turn to the localization properties, and show the nontrivial
effect stemming from the non-Abelian gauge. In order to assess the
localization properties, we first perform the numerical calculations
of the IPRs, which is shown in Fig.~\ref{fig2:IPR}(b).

Obviously, the IPR diagram in Fig.~\ref{fig2:IPR} (b) provides four
regimes separated by $\lambda_{c1}$, $\lambda_{c2}$, and
$\lambda_{c0}$, \textit{i.e.}, pure delocalization phase, coexistence
phase \uppercase\expandafter{\romannumeral1}, and coexistence phase
\uppercase\expandafter{\romannumeral2}, and pure localization phase.
When the modulation strength $\lambda$ is sufficiently small
($<\lambda_{c1}$), IPR values are approximately vanishing for all
states, indicating that all states are extended and the cold atoms
are in the pure delocalization phase. Coexistence phases emerge
within the regime from $\lambda_{c1}$ to $\lambda_{c2}$. The term
``coexistence" herein denotes that the localized and extended states
can be found at the same disorder strength, while they are separated
by critical energies called mobility edges. The two coexistence
phases, labeled by \RomanNumeralCaps{1} and \RomanNumeralCaps{2}, are
separated by the self-dual point $\lambda_{c0}=1$. As $\lambda$ is
increased further ($>\lambda_{c2}$) the non-Abelian system enters
into the pure localization phase and all states are found to be
localized, illustrated by the near-unity IPRs. Apparently, this
diagram is qualitatively different from the A-AAH model, wherein only
two phases of pure delocalization and localization are supported. The
unique coexistence phases arise from the introduction of non-Abelian
gauge.

\begin{figure}[htbp]
\centering
\includegraphics[width=1.0\linewidth]{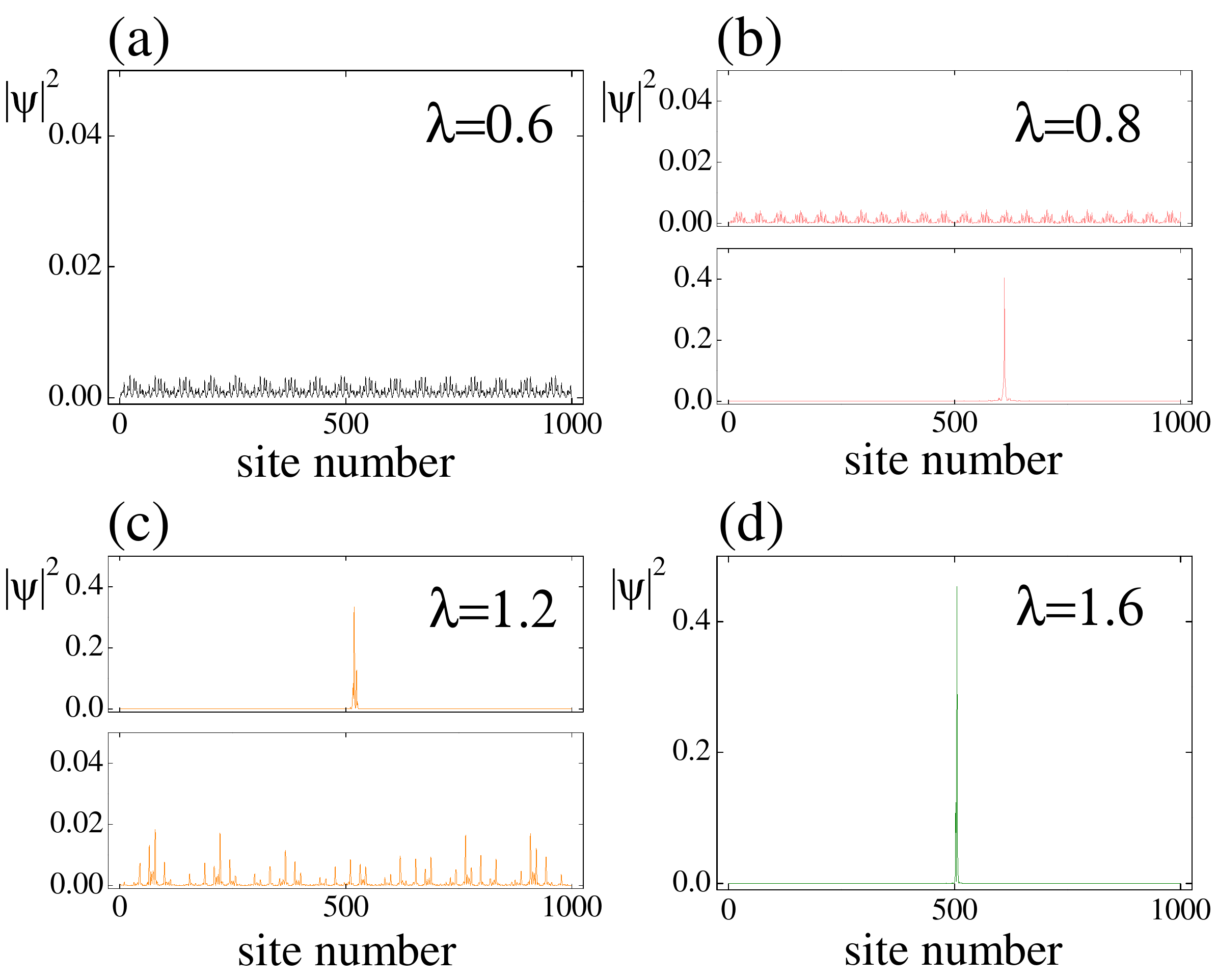}
\caption{(Color online) For a fixed non-Abelian gauge component $q=0.3\pi$,
corresponding to Fig.~\ref{fig2:IPR} (b),
the typical states in various localization regimes, i.e.,
$\lambda=0.6$ (a), $\lambda=0.8$ (b), $\lambda=1.2$ (c), $\lambda=1.6$ (d).}
\label{fig3:states}
\end{figure}

To favor that the distinctions of IPRs do guarantee various phases,
we summarize in Fig.~\ref{fig3:states} the typical states at various
$\lambda$'s. In the purely delocalized regime, all the states extend
over the entire system [Fig.~\ref{fig3:states}(a)]. In the
coexistence phase \uppercase\expandafter{\romannumeral1}, the
extended and exponentially localized states present simultaneously
for the same modulation [Fig.~\ref{fig3:states}(b)]. In the
coexistence phase \uppercase\expandafter{\romannumeral2}
($\lambda_{c2}>\lambda>\lambda_{c0}$) the spatial behavior is similar
to that of phase \uppercase\expandafter{\romannumeral1}
[Fig.~\ref{fig3:states}(c)]. At the strength $\lambda>\lambda_{c2}$,
all the states become spatially localized, as shown in
Fig.~\ref{fig3:states}(d).

In order to further corroborate our findings on the appearance of
coexistence phases in the quasiperiodic non-Abelian optical lattices
(that is so far based on IPR analysis of finite size systems), we now
implement the spectral analysis to identify the phase diagram.
The decomposition of energy spectrum is intimately related to the
localization property of the corresponding
system~\cite{KohmotoPRL1983,KohmotoPRL1989}. Specifically, an
absolutely continuous spectrum corresponds to a delocalization phase.
While a pure point spectrum is associated with the purely localized
regime as in randomly disordered systems. Finally, there is a
critical regime at the metal insulator transition, where the spectrum
becomes singular continuous. Thereby, we can resort to the total
width of allowed energy bands to diagnose various phases.

Note that the quasiperiodic optical lattice is a non-periodic
structure, and therefore it is not possible to access the band
spectrum directly. Hence, we approximate the irrational number
$\Omega=(\sqrt5-1)/2$ by the ratio of two successive Fibonacci terms
$\Omega'=F_{l-1}/F_{l}$, where the Fibonacci sequence is defined by
the recurrence relation $F_{l}=F_{l-1}+F_{l-2}$. As $l$ increases the
rational number $\Omega'$ converges to the golden ratio. Within this
approximation the quasiperiodic model reduces to a periodic
superlattice which has a unit cell of length $F_l$. Thereby, as $l
\to \infty$ the spectrum of the periodic superlattices asymptotically
approaches that of quasiperiodic case and consequently the band width
of quasiperiodic lattices can be accessed. Via its spectral measures,
we can define the decomposition of the spectrum of the NA-AAH
quasiperiodic optical lattices, and further identify the phase
diagram.

\begin{figure}[htbp]
\centering
\includegraphics[width=0.70\linewidth]{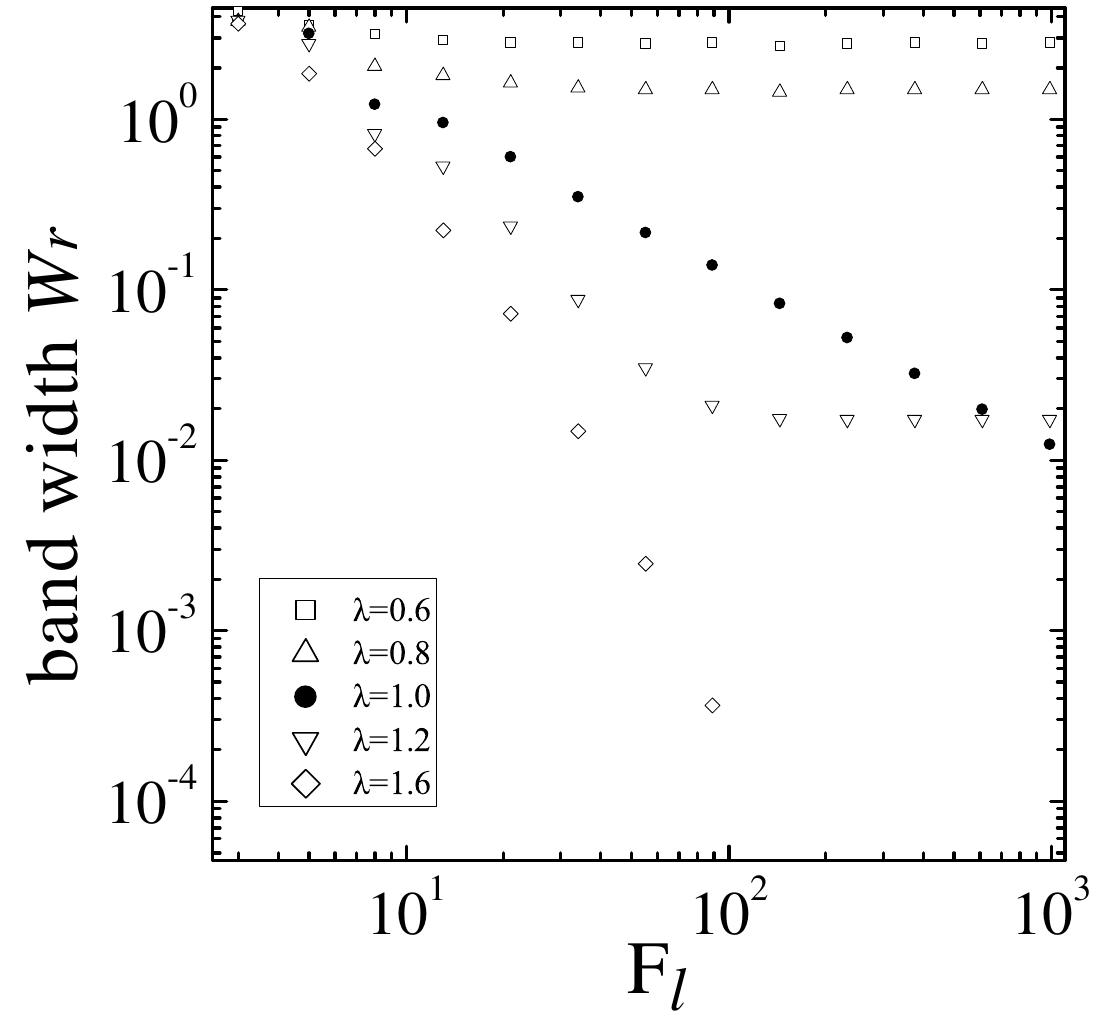}
\caption{(Color online) (a) Total width $\mathcal{W}_r$ of allowed
energy bands versus the period of the
NA-AAH lattice, being equal Fibonacci number
$F_l$. The limit $F_l\rightarrow\infty$ represents the quasiperiodic system.
All data points are with a fixed non-Abelian gauge component
$q=0.3\pi$. Please note the double logarithmic plot.} \label{fig4:bandwidth}
\end{figure}

We first test our method by monitoring the development of the total
band width $\mathcal{W}_r(\lambda,~q)$ of the real-space Hamiltonian
(Eq.~\ref{eq:nonabelianharper}) with the increase of the superlattice
period, \textit{viz.} Fibonacci terms $F_l$, for a given modulation
strength $\lambda$ and non-Abelian gauge $q$. As shown in
Fig.~\ref{fig4:bandwidth}, the band widths exhibit a falling-off,
accompanied by various tendencies dependent on the modulation
strengths. For the weak modulations ($\lambda=0.6,$ $0.8$),
$\mathcal{W}_r$ will converge to a finite value as one follows the
Fibonacci sequence. The $\lambda=1.2$ curve also presents the similar
behavior. This shows the existence of extended states for these
modulations. On the other hand, an algebraic decay $\mathcal{W}_r
\propto F_{(l)}^{-\delta}$ emerges at $\lambda=1$ (solid circles in
Fig.~\ref{fig4:bandwidth}), demonstrating that this is Cantor
spectrum and a transition occurs~\cite{KohmotoPRL1983}. At the
stronger modulation ($\lambda=1.6$), by contrast, $\mathcal{W}_r$
shrinks to zero \emph{exponentially}, signaling a point spectrum
attained. This vanishing asymptotic value ($\mathcal{W}_r \approx 0$)
directly marks the regime of pure localization of the NA-AAH
quasiperiodic lattice. These results are in agreement with the IPR
analysis aforementioned.

Importantly, it should be noted that $\mathcal{W}_r$ alone is
inadequate to distinguish the coexistence phases from the pure
delocalization phases. This is ascribed to the fact that the
finiteness of $\mathcal{W}_r$ is insensitive to the mixing of the
point spectrum (associated with the localized states) inside the
continuous spectrum (associated with the extended states). Hence,
another metric we use is the band width in dual space, \textit{i.e.,}
the band-width $\mathcal{W}^\ast$ of
Hamiltonian~\ref{eq:nonabelianharper-in-k}. This quantity quantizes
the decomposition of the energy spectrum in dual space. Note that the
two quantities $\mathcal{W}_r$ and $\mathcal{W}^\ast$ are
complementary in discerning the localization phases, given the
self-duality between Hamiltonians~\ref{eq:nonabelianharper} and
~\ref{eq:nonabelianharper-in-k}. Thereby, the simultaneous measures
of both quantities can serve as an order parameter of the phase
diagram. To be specific, the presence of the pure metal phase is
characterized by the finite $\mathcal{W}_r$ and the vanishing
$\mathcal{W}^{\ast}$, whereas the pure localization phase is
indicated by the zero $\mathcal{W}_r$ and the finite
$\mathcal{W}^{\ast}$. In between, a coexistence regime is found when
both indices are simultaneously finite. Therefore, the combination of
$\mathcal{W}_r$ and $\mathcal{W}^\ast$ will provide unambiguous
evidence for distinct phases.


\begin{figure}[htbp]
\centering
\includegraphics[width=0.70\linewidth]{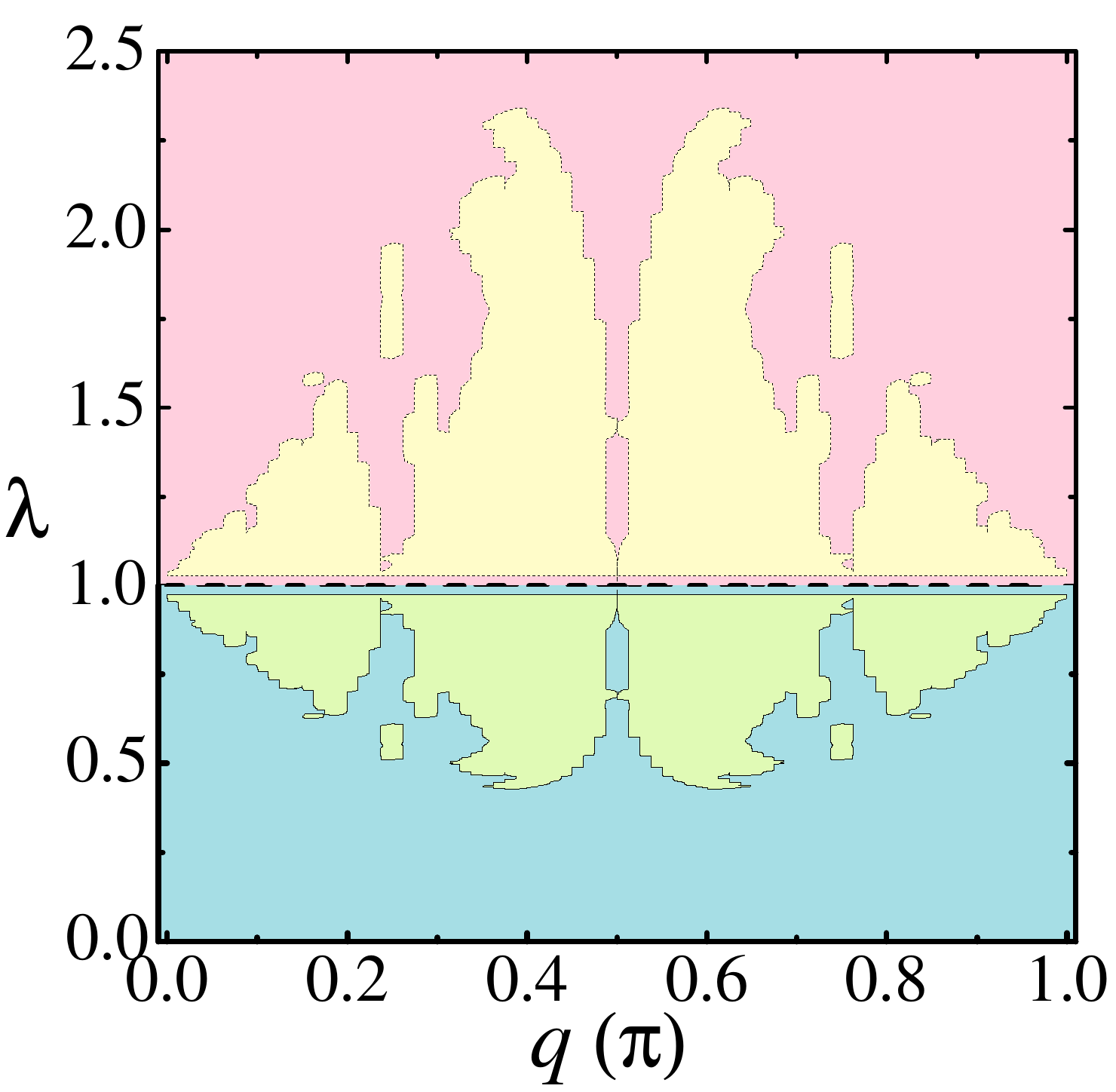}
\caption{(Color online) Localization phase diagram of the quasiperiodic
NA-AAH lattice obtained from the band width data $\mathcal{W}_r$ and $\mathcal{W}^{\ast}$.
Solid lines are the separatrices to divide different regions,
which are extracted through the vanishing values of $\mathcal{W}_r$ and
$\mathcal{W}^{\ast}$. Due to the computational
accuracy we take $1.0\times 10^{-7}$.
Clearly four regions are observed. The blue-shaded region is characterized
by $\mathcal{W}_r>1.0\times 10^{-7}$ and $\mathcal{W}^{\ast} < 1.0\times 10^{-7}$.
In the yellow-shaded region both $\mathcal{W}_r$
and $\mathcal{W}^{\ast}$ are simultaneously larger than
$1.0\times 10^{-7}$. This area is further separated into two parts by the self-dual
line $\lambda_{c0}=1$ (dashed line). The red-shaded region
is with $\mathcal{W}_r<1.0\times 10^{-7}$ and $\mathcal{W}^{\ast} > 1.0\times 10^{-7}$.
} \label{fig5:phase-diagrm}
\end{figure}

Based on these arguments, we scan the band width data to extract the
separatrices among different phases in $\lambda-q$ plane. As seen
from Fig.~\ref{fig5:phase-diagrm}, we find that for a non-Abelian
gauge $q$, the quasiperiodic NA-AAH lattice passes through four
phases, including the pure delocalization-coexistence
\uppercase\expandafter{\romannumeral1}-coexistence
\uppercase\expandafter{\romannumeral2}-pure localization, as
$\lambda$ increases. In the blue-shaded area, we always find a finite
value of $\mathcal{W}_r$ ($\mathcal{W}_r>0$) whereas a vanishing
$\mathcal{W}^{\ast}$ ($\mathcal{W}^{\ast} \approx 0$). This
characterizes the absence of any localized state, namely the pure
delocalization regime. In the red-shaded area, we find a fully
localized phase, which is marked by $\mathcal{W}_r \approx 0$ while
$\mathcal{W}^\ast>0$. This is the pure localization regime. In
between, an intermediate regime (yellow-shaded area) is found where
$\mathcal{W}_r>0$ and $\mathcal{W}^{\ast}
>0$. This directly illustrates the cooccurrence
of extended and localized states in this area, in which mobility
edges are present. Furthermore, the coexistence phase is split into
two distinct regions \RomanNumeralCaps{1} and \RomanNumeralCaps{2} by
the self-dual point $\lambda_{c0}=1$. It is worth noting that for
$q=0, ~\pi/2,~\mathrm{and}~\pi$ the coexistence phases essentially
disappear. In other words, the localization transition is similar to
that of Abelian AAH lattices. The reason lies on the fact that the
NA-AAH lattice with $q=0,~ \pi$ can be mapped accurately onto two
decoupled Abelian replicas which have no mobility edges individually.
And $q=\pi/2$ reduces Eq.~\ref{eqn:NA_Hofstadter_ham} to the Abelian
flux model~\cite{GoldmanPRL2009}. Moreover, we find the reentrant
coexistence phases appears far away from the self-dual line, as
indicated by the circled areas in the pure delocalization
(localization) phase. Interestingly, we notice that
Fig.~\ref{fig5:phase-diagrm} embodies beautiful symmetries. Firstly,
the localization phase diagram is symmetric with respect to the axis
$q=\pi/2$. We also find the reflection symmetry about $\lambda=1$,
inversely scaled by $1/\lambda$. This $\lambda\leftrightarrow
\frac{1}{\lambda}$ symmetry results from the self-duality of the
NA-AAH Hamiltonian.

\section{Discussion and conclusion}
\label{sec:conclusions}

Before concluding we should point that the ultracold atoms studied
here are immersed in a non-Abelian gauge. These are not the
non-Abelian particles respecting the non-Abelian exchange statistics.
Therefore, our work differs fundamentally from the previous
work~\cite{AAHpwave}. Therein are the localization of $p$-wave. On
the other hand, there has been one work proposed a quasiperiodic
lattices with non-Abelian gauge~\cite{SatijaPRL2006}. However, the
non-Abelian gauge used there is characterized by the spatial
inhomogeneity, and the Hamiltonian has not yet been determined to be
self-dual.

In conclusion, we have combined two vital concepts, disorder and
non-Abelian gauge, and unveiled rich localization phases of cold
atoms in the non-Abelian quasiperiodic optical lattices. Besides the
pure metal and pure insulator, we found the emergent coexistence
phases. In other words, for a quasiperiodic optical lattice, the
non-Abelian gauge can drive a localization transition in an
engineered manner. This represents the high sensitivity of global
transport of ultracold atoms to the non-Abelian configurations.
Therefore, our results are fundamentally different from the
extensively studied Abelian quasiperiodic problems. We also developed
a method to identify the localization phase diagram: diagnosing the
decomposition of the energy spectra in both real and dual spaces.
This method is accurate and general. In a word, the marriage of
Anderson localization and non-Abelian gauge potential may open up
exciting new physics and could enrich our understanding of the
metal-insulator transition.

\begin{acknowledgements}
The authors are indebted to Prof. Ray Kuang Lee from National Tsing
Hua University for stimulating discussions. This work is supported by
the National Natural Science Foundation of China (11604231); Natural
Science Foundation of Jiangsu Province under Grant (BK20160303);
Natural Science Foundation of the Jiangsu Higher Education
Institutions of China (16KJB140012); Priority Academic Program
Development (PAPD) of Jiangsu Higher Education Institutions.
\end{acknowledgements}

\bibliographystyle{apsrev4-1}

\begin{references}

\bibitem{AB} Y. Aharonov and D. Bohm, ``Significance of
    electromagnetic potentials in the quantum theory," Phys. Rev. {\bf115}, 485 (1959).

\bibitem{GalitskiPhysTod2019} V. Galitski, Ian Spielman, and G.
    Juzeli\={u}nas, ``Artifical Gauge Fields with Ultracold Atoms,"
    Physics Today {\bf72}, 39 (2019).

\bibitem{JakschNJP2003} D. Jaksch and P. Zoller, ``Creation of
    effective magnetic fields in optical lattices: the Hofstadter
    butterfly for cold neutral atoms," New J.
    Phys. {\bf 5}, 56 (2003).

\bibitem{NonAbelian:20051} K. Osterloh, M. Baig, L. Santos, P.
    Zoller, and M. Lewenstein, ``Cold Atoms in Non-Abelian Gauge
    Potentials: From the Hofstadter Moth to Lattice Gauge," Phys. Rev.
  Lett. {\bf 95}, 010403 (2005).

\bibitem{NonAbelian:20052} J. Ruseckas, G. Juzeli\={u}nas, P.
    \"{O}hberg, and M. Fleischhauer, ``Non-Abelian Gauge Potentials
    for Ultracold Atoms with Degenerate Dark States," Phys. Rev.
  Lett. {\bf 95}, 010404 (2005).

\bibitem{DalibardRMP2011} Jean Dalibard, Fabrice Gerbier, Gediminas
    Juzeli\={u}nas, and Patrik \"{O}hberg, ``Artificial gauge
    potentials for neutral atoms," Rev. Mod. Phys. {\bf83}, 1523
    (2011).

\bibitem{GoldmanRPP2014} N. Goldman, G. Juzeli\={u}nas, P.
    \"{O}hberg, I. B. Spielman, ``Light-induced gauge fields for ultracold atoms,"
    Rep. Prog. Phys. {\bf77}, 126401 (2014).

\bibitem{AidelsburgerCRP2018} M. Aidelsburger, S. Nascimbene, and N.
    Goldman, ``Artificial gauge fields in materials and
    engineered systems," C. R. Phys. {\bf 19}, 394 (2018).

\bibitem{WilczekPRL1984} F. Wilczek and A. Zee, ``Appearance of Gauge
    Structure in Simple Dynamical Systems," Phys. Rev. Lett. {\bf52}, 2111 (1984).


\bibitem{AndersonPR1958} P. W. Anderson, ``Absence of Diffusion in
    Certain Random Lattices," Phys. Rev. {\bf 109},
    1492 (1958).

\bibitem{GangofFour} E. Abrahams, P. W. Anderson, D. C. Licciardello,
    and T. V. Ramakrishnan, ``Scaling Theory of Localization: Absence
    of Quantum Diffusion in Two Dimensions," Phys. Rev. Lett. {\bf 42}, 673 (1979).

\bibitem{LeeRMP1985} P. A. Lee and T. V. Ramakrishnan, ``Disordered
    electronic systems," Rev. Mod.
    Phys. {\bf 57}, 287 (1985).

\bibitem{Kramer1993} B. Kramer and A. MacKinnon, ``Localization
    theory and experiment," Rep. Prog. Phys.
    {\bf 56}, 1469 (1993).

\bibitem{MirlinRMP2008} F. Evers and A. D. Mirlin, ``Anderson
    transitions,"  Rev. Mod. Phys. {\bf 80}, 1355 (2008).

\bibitem{BlochRMP2008} I. Bloch, J. Dalibard, and W. Zwerger,
    ``Many-body physics with ultracold gases," Rev. Mod. Phy. {\bf80}, 885
    (2008).

\bibitem{CBS1} F. Jendrzejewski \textit{et al}., ``Coherent
    Backscattering of Ultracold Atoms," Phys. Rev. Lett. {\bf 109}, 195302 (2012).

\bibitem{CBS2} K. M\"{u}ller \textit{et al}., ``Suppression and
    Revival of Weak
    Localization through Control
    of Time-Reversal Symmetry,"  Phys. Rev. Lett. {\bf114},
    205301 (2015).

\bibitem{BillyNature2008} J. Billy \textit{et al}., ``Direct
    observation of Anderson localization of matter waves in a
    controlled disorder ," Nature {\bf 453},
    891 (2008).

\bibitem{RoatiNature2008} G. Roati \textit{et al}., ``Anderson
    localization of a non-interacting Bose Einstein condensate,"
    Nature {\bf 453}, 895 (2008).

\bibitem{3DAL1} S. S. Kondov, W. R. McGehee, J. J. Zirbel, and B.
    DeMarco, ``Three-Dimensional Anderson Localization of Ultracold Matter," Science
    {\bf 334}, 66 (2011).

\bibitem{3DAL2} F. Jendrzejewski \textit{et al}.,
    ``Three-dimensional localization of ultracold atoms in an optical disordered potential,"
    Nature Phys. {\bf8}, 398 (2012).

\bibitem{me1} G. Semeghini\textit{et al}., ``Measurement of the
    mobility edge for 3D Anderson localization," Nat. Phys. {\bf11}, 554 (2015).

\bibitem{AA} S. Aubry and G. Andr\'{e}, ``Analyticity breaking and
    Anderson localization in incommensurate lattices," Ann. Isr.
    Phys. Soc. {\bf 3}, 133 (1980).

\bibitem{Harper} P. G. Harper, ``Single Band Motion of Conduction
    Electrons in a Uniform Magnetic Field," Proc. Phys. Soc. A {\bf
    68}, 874 (1955).

\bibitem{Sarma88} S. Das Sarma, S. He, and X. C. Xie, ``Mobility Edge
    in a Model One-Dimensional Potential," Phys.Rev. Lett. {\bf61}, 2144
    (1988).

\bibitem{Thouless88} D. J. Thouless, ``Localization by a Potential
    with Slowly Varying Period," Phys. Rev. Lett. {\bf61}, 2141 (1988).

\bibitem{Holthaus2007} Dave J. Boers, Benjamin Goedeke, Dennis
    Hinrichs, and Martin Holthaus, ``Mobility edges in bichromatic optical
    lattices," Phys. Rev. A. {\bf75}, 063404 (2007).

\bibitem{Biddle10} J. Biddle and S. Das Sarma, ``Predicted Mobility
    Edges in One-Dimensional Incommensurate Optical Lattices: An Exactly
    Solvable Model of Anderson Localization," Phys. Rev. Lett. {\bf104},
    070601 (2010).

\bibitem{SF1} S. Ganeshan, J. H. Pixley, and S. Das Sarma, ``Nearest
    Neighbor Tight Binding Models with an Exact Mobility Edge in One
    Dimension," Phys. Rev. Lett. {\bf114}, 146601 (2015).

\bibitem{SF2} M. L. Sun, G. Wang, N. B. Li, and T. Nakayama,
    ``Localization-delocalization transition in self-dual quasiperiodic
    lattices," Europhys. Lett. {\bf110}, 57003 (2015).

\bibitem{SF3} M. Johansson
    and R. Riklund, ``Self-dual model for one dimensional incommensurate
    crystals including next-nearest-neighbor hopping, and its
    relation to the Hofstadter model," Phys. Rev. B {\bf43}, 13468 (1991).

\bibitem{SF4} S. Gopalakrishnan, ``Self-dual quasiperiodic systems
    with power-law hopping," Phys. Rev. B {\bf96}, 054202 (2017).

\bibitem{sarma4} X. Li, J. H. Pixley, D. L. Deng, S. Ganeshan, and
    S. Das Sarma, ``Quantum nonergodicity and fermion localization in
    a system with a single-particle mobility edge," Phys. Rev. B {\bf 93}, 184204 (2016).
\bibitem{sarma5} X. Li, X. Li and S. Das Sarma, ``Mobility edges in
    one-dimensional bichromatic incommensurate potentials,"
    Phys. Rev. B {\bf96}, 085119 (2017).

\bibitem{selfdual4} L. Gong, Y. Feng, and Y. Ding, ``Anderson
    localization in one-dimensional quasiperiodic lattice models with
    nearest- and next-nearest-neighbor hopping," Phys.
    Lett. A {\bf 381}, 588 (2017).

\bibitem{me2} H. P. L\"{u}schen \textit{et al}., ``Single-Particle
    Mobility Edge in a One-Dimensional Quasiperiodic Optical Lattice,"
    Phys. Rev. Lett. {\bf120}, 160404 (2018).

\bibitem{me3} F. A. An, E. J. Meier, and B. Gadway, ``Engineering a
    Flux-Dependent Mobility Edge in Disordered Zigzag Chains," Phys.
    Rev. X
    {\bf8}, 031045 (2018).

\bibitem{HofstadterPRB} D. R. Hofstadter, ``Energy levels and wave
    functions of Bloch electrons in rational and irrational magnetic
    fields," Phys. Rev. B {\bf 14}, 2239 (1976).

\bibitem{GoldmanPRL2009} N. Goldman, A. Kubasiak, A. Bermudez, P.
    Gaspard, M. Lewenstein, and M. A. Martin-Delgado, ``Non-Abelian Optical Lattices:
    Anomalous Quantum Hall Effect and Dirac Fermions," Phys. Rev. Lett. {\bf103}, 035301
    (2009).

\bibitem{AHE2} N. Goldman, A. Kubasiak, P. Gaspard, and M.
    Lewenstein, ``Ultracold atomic gases in non-Abelian gauge potentials:
    The case of constant Wilson loop," Phys. Rev. A {\bf 79}, 023624 (2009).

\bibitem{AHE3} J. M. Hou, W. X. Yang, and X. J. Liu, ``Massless Dirac
    fermions in a square optical lattice," Phys. Rev.
    A {\bf 79}, 043621 (2009).

\bibitem{AE} A. Bermudez \textit{et al.}, ``Wilson Fermions and Axion
    Electrodynamics in Optical Lattices," Phys. Rev. Lett.
    {\bf 105}, 190404 (2010).

\bibitem{KohmotoPRL1983} M. Kohmoto, ``Metal-Insulator Transition and
    Scaling for Incommensurate Systems," Phys. Rev. Lett. {\bf51}, 1198 (1983).
\bibitem{KohmotoPRL1989} H. Hiramoto and M. Kohmoto,
    ``New Localization in a Quasiperiodic System,"
    Phys. Rev. Lett. {\bf62}, 2714 (1989).

\bibitem{AAHpwave} Jun Wang, Xia-Ji Liu, Gao Xianlong, and Hui Hu,
    ``Phase diagram of a non-Abelian Aubry-Andr\'{e}-Harper model with
    \textit{p}-wave superfluidity," Phys. Rev. B {\bf93}, 104504 (2016).

\bibitem{SatijaPRL2006} I. Satija, D. Dakin, and C. Clark,
    ``Metal-Insulator Transition Revisited for Cold Atoms in
    Non-Abelian Gauge Potentials," Phys Rev. Lett. {\bf97}, 216401 (2006).
\end{references}

\end{document}